\documentclass[9pt,twocolumn,twoside]{opticajnl}
\journal{opticajournal} 

\setboolean{shortarticle}{false}

\usepackage{amsmath,commath,xfrac}
\usepackage{physics}
\usepackage{bm}

\usepackage{lineno} 

\usepackage{siunitx}
\newcommand{\SIadj}[2]{\SI[number-unit-product={\text{-}}]{#1}{#2}}
\DeclareSIUnit\bar{bar}

\newcommand{\etal}[1]{\mbox{{#1} \textit{et al.}}}
\newcommand{\ie}{\textit{i}.\textit{e}.,\ }
\newcommand{\eg}{\textit{e}.\textit{g}.,\ }
\newcommand*{\Exp}{\mathrm{e}}
\newcommand*\diff{\mathop{}\!\mathrm{d}}

\newcommand{\RN}[1]{%
\textup{\uppercase\expandafter{\romannumeral#1}}%
} 
\newcommand{\rn}[1]{%
\textup{\lowercase\expandafter{\romannumeral#1}}%
} 

\usepackage[version=4]{mhchem} 

\title{Analysis and control of Raman phonon dynamics for enhanced optical frequency conversion}

\author[1]{Yi-Hao Chen}
\author[1]{Frank Wise}

\affil[1]{School of Applied and Engineering Physics, Cornell University, Ithaca, New York 14853, USA}

\affil[*]{yc2368@cornell.edu}

\begin{abstract}
Raman phonons are quantized molecular motions that arise from the inelastic scattering of light and mediate a wide range of spectroscopic and nonlinear optical phenomena. These can play a major role in frequency-conversion processes, but commonly-used theoretical treatments based on the Raman gain spectrum largely neglect the phonons and their dynamical interaction with the field. In this work, we clarify the physical role of Raman phonons within a recently-developed time-domain framework based on the Raman-induced index modulation ${\triangle\epsilon_R(t)}$, and show that phonons correspond to the oscillatory component of the Raman-induced index modulation. The analysis further reveals a linear phonon-mediated interaction embedded within Raman scattering, in which optical fields couple through wave-vector matching with existing phonons. This mechanism underlies, but has been neglected in, coherent Stokes and anti-Stokes scattering, as well as molecular modulation. Building on this insight, we introduce a phonon-controlled approach that enables efficient conversion into a selected Stokes order by tuning the wave-vector-matching relation between the driven phonons and the targeted Raman process, and we confirm the approach by numerical simulations that consider realistic Raman dynamics. These results provide a clearer physical interpretation of Raman phonons and their dynamics, and offer new strategies for controlling Raman interactions.
\end{abstract}

\setboolean{displaycopyright}{false} 

\begin{document}
\maketitle

\section{Introduction}
Raman phonons refer to the quantized motions in matter that originate from the inelastic Raman scattering of light \cite{Smekal1923,Raman1928,Raman1928a,Raman1928b,Landsherg1928b,Landsherg1928a,Landsherg1928}. The scattered light is redshifted through exchange of energy between photons and internal degrees of freedom of the medium, driving molecular motions, and thus creating phonons. This Raman phonon excitation provides direct access to the microscopic dynamics of matter and forms the basis of a wide range of spectroscopic \cite{Ploetz2007,Min2025} and nonlinear optical phenomena \cite{Hill1976,Boyraz2004,Sirleto2020}.

Traditional approaches to the analysis of Raman scattering isolate the oscillatory component of the material response by solving the coupled phonon equation with an assumed $\Exp^{i\left(\beta^{\text{ph}}z-\omega_Rt\right)}$ wave solution [$\beta^{\text{ph}}$: phonon wave vector; $\omega_R$: phonon oscillation angular frequency which is equal to the Raman transition angular frequency] \cite{Bloembergen1964,Raymer1990,Boyd2008}. This has been quite successful in explaining a wide range of properties of Raman scattering, and has been employed in diverse spectroscopic and frequency-shifting applications of Raman scattering.

\begin{figure}[!ht]
\centering
\includegraphics[width=\linewidth]{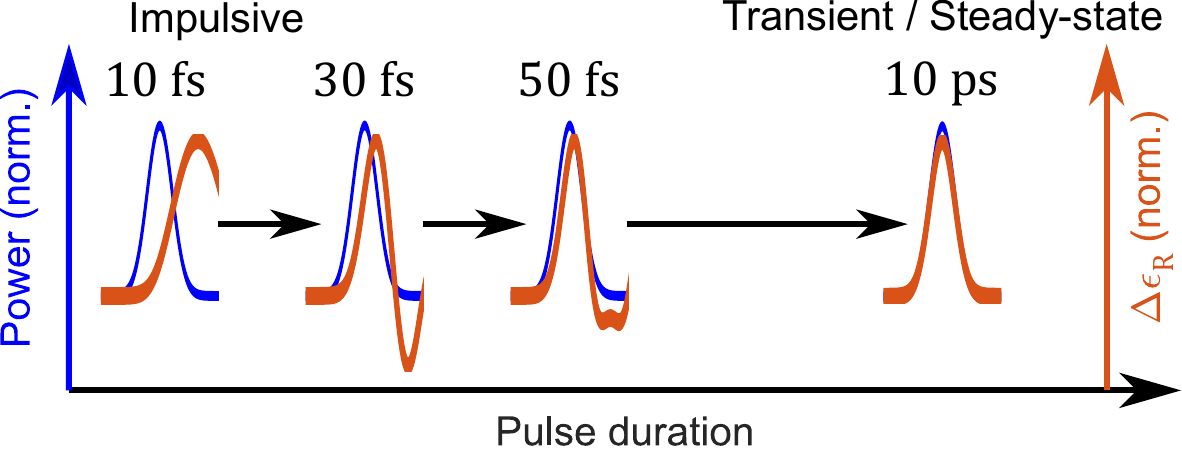}
\caption{Raman-induced index modulation driven by a pulse of different durations \cite{Chen2024}. The examples here use \SI{56.8}{\fs} as the Raman period, which corresponds to the $S(1)$ rotational Raman transition in \ce{H2}. Impulsive regime: $\tau_0\ll\tau_R,T_2$; transient regime: $\tau_0\ll T_2,\tau_0\gg\tau_R$; steady-state regime: $\tau_0\gg\tau_R,T_2$. $\tau_0$: pulse duration, $\tau_R=1/\nu_R$: Raman period, $\nu_R=\omega_R/(2\pi)$: Raman transition frequency, $T_2=1/\gamma_2$: dephasing time, $\gamma_2$: dephasing rate.}
\label{fig:Index_change_regimes}
\end{figure}

Recently, we have developed a new perspective for understanding Raman dynamics based on the temporally-varying Raman-induced index modulation ${\triangle\epsilon_R(z,t)}\propto{R(t)\ast\abs{A(z,t)}^2}$ (Fig.~\ref{fig:Index_change_regimes}) \cite{Chen2024,Chen2026}, where $R(t)$ is the Raman temporal response and $A(z,t)$ is the optical field assuming an unchanged spatial profile, \eg in an optical fiber. The nonlinear Raman field increment follows $\left[\partial_zA(z,t)\right]_{\mathcal{R}}\propto {i\left[\triangle\epsilon_R(z,t)A(z,t)\right]}$. In contrast to the traditional approach, this framework enables direct visualization of the full spectrotemporal Raman dynamics, which combine frequency-dependent Raman gain and time-dependent phonon dynamics. In the long-pulse regime, the pulse-following component of $\triangle\epsilon_R(z,t)$ contributes to Raman-enhanced self-phase modulation (SPM), facilitating supercontinuum generation and ultrashort-pulse formation \cite{Belli2015,Hosseini2018,Konyashchenko2019,Beetar2020}. In the ultrashort-pulse regime, the pulse impulsively drives the Raman scattering and inertia induces a significantly-delayed index modulation so that the pulse interacts with only the initial rising edge of the index wave. Operation in the intermediate regime combines features of both limits, yielding enhanced spectral broadening accompanied by a net redshift \cite{Fan2020,Carpeggiani2020,Eisenberg2025}. Moreover, the slow inertial response smooths the nonlinear index profile, facilitating high-fidelity spectral compression \cite{Wang2026a}. Although this framework provides a clear and comprehensive understanding of Raman dynamics across different temporal regimes, it leads to a new question regarding the connection between the index modulation and the traditional concept of quantized phonons: Is the pulse-following index that originates from Raman scattering a part of the Raman phonon? This article will explain the role of the Raman phonon in terms of index modulation.

In addition to providing a fresh time-domain viewpoint, this framework reveals the linear phonon-mediated process embedded within Raman scattering. In the case of two pulses incident on a Raman-active medium, phonons excited by the first pulse scatter the second pulse. Depending on the wave-vector mismatch between the pre-existing phonons and the Raman process driven by the second pulse, the scattering can produce discrete Stokes \cite{Konyashchenko2007,Konyashchenko2008,Didenko2015,Vicario2016,Konyashchenko2017} or anti-Stokes \cite{Bauerschmidt2015a} radiation, corresponding to phonon generation or annihilation, respectively. Although this mechanism has been demonstrated experimentally, a valid theoretical analysis has only recently appeared \cite{Chen2023,Chen2024}, and even then the discussion remained limited to two-pulse scenarios. Here, we further recognize that this linear phonon-mediated process constitutes the underlying origin of coherent Stokes and anti-Stokes Raman scattering (CSRS/CARS) \cite{Nandakumar2009,Freudiger2008,Ozeki2009,Hamer2024,Maker1965,Levenson1972,Duncan1982,Zumbusch1999}, whose well-known phase-matching condition originates from the phonon-mediated wave-vector matching. Moreover, this linear mechanism is distinct from the traditional interpretation based on nonlinear Raman (exponential) amplification. Due to the limitations of the traditional steady-state treatment, the linear mechanism has been overlooked. Notably, elucidating the underlying phonon interaction enables control of Raman generation beyond current capabilities. As a first example, we introduce a phonon-controlled technique that suppresses early-stage cascaded Raman generation. By deliberately manipulating the light-phonon interaction, phonons generated by the first-order Stokes process exert minimal influence on subsequent Stokes generation, which effectively confines the energy to a specific order and enhances the frequency-conversion efficiency. Although the results presented here are theoretical, straightforward experimental realization of the approach in gas-filled hollow-core fiber is illustrated.

Given the breadth of Raman-scattering research accumulated over nearly a century, it is natural to ask what genuinely new physical insights remain to be uncovered. The framework developed here addresses this question by revealing aspects of Raman dynamics, and of CSRS/CARS in particular, that have been overlooked or misinterpreted in traditional treatments. Readers already familiar with CSRS/CARS may therefore wish to consult Sec.~\ref{sec:Comparison}, where we summarize the novel aspects of this work and clarify how they differ from existing analyses.

\section{Raman phonons}\label{sec:Raman_phonons}
We start by briefly reviewing and summarizing the understanding of Raman phonons. Optical phonons have been well studied in crystalline solids using periodic Bloch-wave descriptions \cite{Ashcroft2022}, where the relative motion of atoms within the basis gives rise to high-frequency vibrational modes that couple strongly to electromagnetic fields. These modes play a central role in infrared absorption, Raman scattering, and a wide range of light-matter interactions, owing to their ability to modulate the crystal's polarization at optical frequencies.

\begin{figure}[!ht]
\centering
\includegraphics[width=\linewidth]{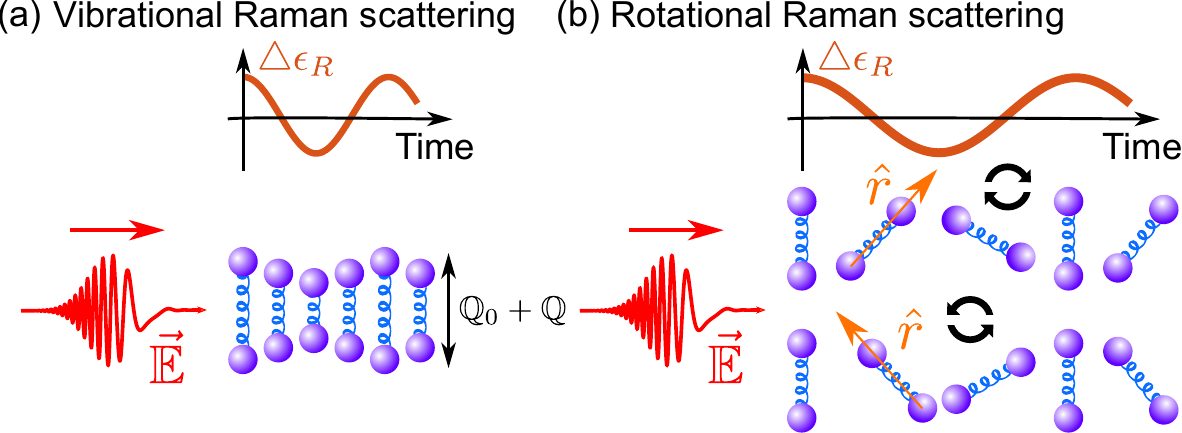}
\caption{Raman scattering due to (a) vibrational or (b) rotational molecular motions. A linearly-polarized field is chosen for illustration here.}
\label{fig:vib_rot_models}
\end{figure}

The concept of optical phonon can be generalized beyond crystalline solids in the context of Raman scattering. The characteristic \emph{out-of-phase} motion of atoms in a lattice provides a natural analogy to the vibrational dynamics of isolated molecules, allowing the optical-phonon picture to extend to liquids and gases, where no long-range periodic structure exists (see Supplementary Fig.~S3 in \cite{Chen2024}). In fact, the extension requires the ensemble-averaging over microscopic motions, and the collective oscillations of phonons refer to the macroscopic motions experienced by the optical field. This explains why a rotational Stokes field remains linearly polarized when driven by a linearly-polarized pump field \cite{Gao2022,Eisenberg2025a,Eisenberg2025}. A linearly-polarized pump excites molecular rotations microscopically equally in the clockwise and counterclockwise directions, producing a macroscopic index modulation [Fig.~\ref{fig:vib_rot_models}(b)] that generates a linearly-polarized Stokes field. In contrast, a driving field that is not linearly-polarized preferentially excites rotational Raman scattering in one direction, leading to a vectorial response in the resulting Stokes generation \cite{Chen2024}.

The internal motions of fluids involve vibrations and rotations of molecules, as well as intermolecular interactions. In gases, intermolecular interactions are typically weak and can often be neglected. Conventional descriptions of vibrational and rotational Raman dynamics, therefore, rely on individual molecular vibrations modeled as harmonic oscillators \cite{Boyd2008,Korn1998,Nazarkin1998,Nazarkin1999} and molecular rotations modeled in terms of quantized rotational states \cite{Chen2024,Martin1988,Nibbering1997,Chen2007,Hoque2011,Langevin2019}, respectively (Fig.~\ref{fig:vib_rot_models}). The characteristic motions in these two models are indicated by the intramolecular displacement $\mathbb{Q}$ or molecular alignment angle $\theta$ between the light polarization (vector) $\hat{\mathbb{E}}$ and molecular axis $\hat{r}$. A more general framework employs the quantum-mechanical density matrix $\boldsymbol{\rho}$ \cite{Boyd2008a} to describe the molecular ensemble of both motions, and Raman dynamics in perturbative Raman operations, where the population remains predominantly in the ground state, are governed primarily by the off-diagonal coherence waves $\rho_{ab}$ ($a\neq b$) \cite{Chen2024}:
\begin{equation}
\triangle\epsilon_R(z,t)\propto\sum\Re\left[\rho_{ab}(z,t)\right]
\end{equation}
Connections between these distinct models can be established as follows \cite{Raymer1990,Chen2024}:
\begin{subequations}
\begin{align}
\mathbb{Q} & =\frac{1}{2}\left[\sqrt{\frac{2\hbar}{\mu\omega_{01}}}\rho_{01}^*+\text{c.c.}\right] \label{eq:vib_rho} \\
\rho_{ab} & =\frac{i}{\hbar}\left(\rho^{(0)}_b-\rho^{(0)}_a\right)\frac{\triangle\alpha}{2} \nonumber \\
& \hspace{1em}\times\int_{-\infty}^t\left[\vec{\mathbb{E}}(\tau)\cdot\hat{r}\hat{r}\cdot\vec{\mathbb{E}}(\tau)\right]_{ab}\Exp^{\left(\gamma_{ab}+i\omega_{ab}\right)\left(\tau-t\right)}\diff\tau, \label{eq:rot_rho}
\end{align}
\end{subequations}
with $\mu$ the reduced mass of a diatomic molecule, $\omega_{01}=\omega_{R_{\text{vib}}}$ the vibrational transition angular frequency, $\rho_a^{(0)}$ the unperturbed population at level $a$, $\triangle\alpha$ the polarizability anisotropy, $\vec{\mathbb{E}}$ the real-valued electric field (with complex-valued $A(z,t)$ representing its analytic signal \cite{Chen2025}), and $\gamma_{ab}$ and $\omega_{ab}$ the dephasing rate and the transition angular frequency between rotational levels $a$ and $b$, respectively. The coherence $\rho_{ab}$ corresponds to the \emph{negative-frequency} component of the displacement $\mathbb{Q}$ in vibrations, or directly proportional to the orientation $\left(\vec{\mathbb{E}}\cdot\hat{r}\hat{r}\cdot\vec{\mathbb{E}}\right)$ in rotations, which, for example under linear polarization, becomes $\abs{\vec{\mathbb{E}}}^2\cos^2\left(\sfrac{\pi}{2}-\theta\right)$.

\begin{figure}[!ht]
\centering
\includegraphics[width=\linewidth]{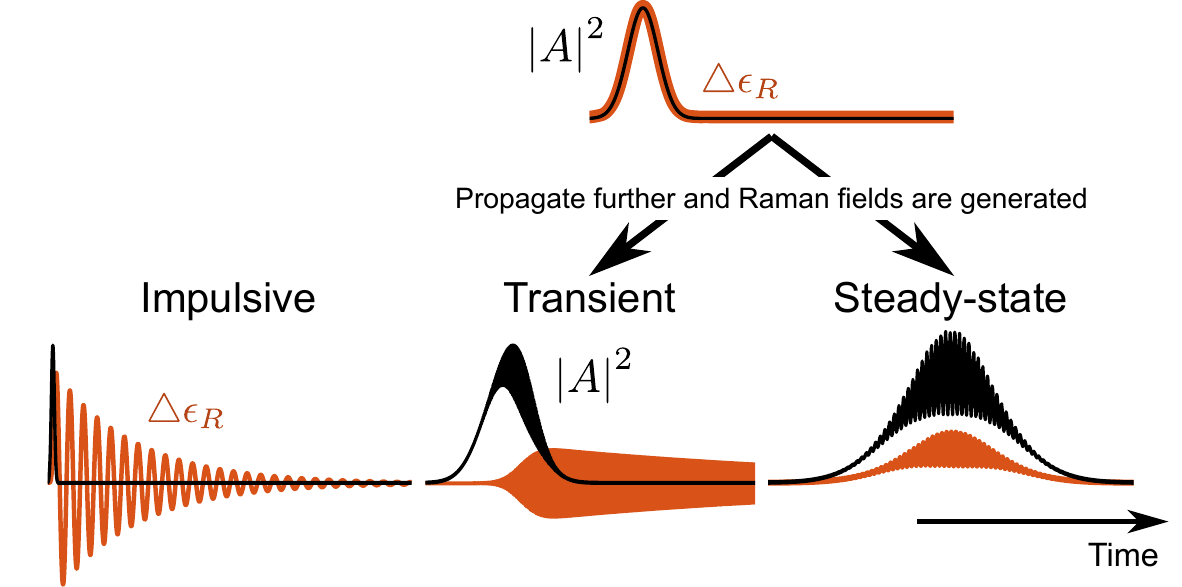}
\caption{Oscillatory Raman-induced index modulation $\triangle\epsilon_R^{\text{osc}}$ in different temporal Raman regimes. Here, $A(z,t)$ is the ``total'' electric field that can exhibit oscillations due to beating between the pump and the Raman fields. In the long-pulse regimes, the oscillatory part in the index can be much stronger than the pulse-following part.}
\label{fig:phonon_vs_index}
\end{figure}

There has been ambiguity in the literature regarding the distinction between the Raman phonon and the coherence (index) wave that involves a pulse-following non-oscillatory part (Fig.~\ref{fig:Index_change_regimes}) \cite{Bauerschmidt2015,Hosseini2017,Loranger2020,Tyumenev2022}. The phonon provides a quantum-mechanical description of the oscillatory components of Raman-induced index modulation (Fig.~\ref{fig:phonon_vs_index}), namely those oscillating around $\omega_R$, which we denote hereafter as $\triangle\epsilon_R^{\text{osc}}$ (see Supplementary Sec.~2 for more discussion of this phonon part of the index in the time-domain picture). In the impulsive regime, the induced index modulation exhibits a pronounced oscillatory behavior that persists beyond the duration of the driving pulse, a response that intuitively corresponds to the excitation of phonons. By contrast, in the long-pulse transient and steady-state regimes, the index predominantly follows the temporal envelope of the optical pulse and does not exhibit intrinsic oscillations. A pulse-following index does not induce an overall redshift for the pump, but enhances spectrally-symmetric SPM, no net energy is transferred to the medium and thus no phonon is generated. Oscillatory behavior emerges only after Raman fields are generated (from vacuum fluctuations or an external weak seed), which interfere to produce high-frequency beating that impulsively drives the oscillatory index response. The amplitude of these oscillations is determined by the strength of the optical beating, which in turn depends on the Stokes-field amplitude and its photon number. As one Stokes photon generated corresponds to one phonon generated, the magnitude of the oscillatory response reflects the number of phonons generated in the Raman process.

\section{Linear phonon-mediated process}
Traditional frameworks of Raman scattering focus primarily on the evolution of the optical fields, while the coupled phonon equation for vibrational $\mathbb{Q}$ \cite{Raymer1990,Bauerschmidt2015a,Bloembergen1964,Bloembergen1967,Raymer1981,Mostowski1981} or rotational $\theta$ \cite{Martin1988,Nibbering1997,Chen2007,Langevin2019} excitation is treated more as an auxiliary mathematical artifice. A clear and physically-intuitive picture is obscured by having a coupled equation. In this article, we analyze Raman phenomena through the oscillatory Raman-induced index modulation $\triangle\epsilon_R(z,t)$ that the field directly interacts with \cite{Chen2024,Chen2026}, which makes the role of the phonon explicit and physically transparent. Within this framework, we describe in detail the interaction between phonons and optical fields, and reveal the overlooked linear phonon-mediated process underlying Raman scattering.

\begin{figure}[!ht]
\centering
\includegraphics[width=\linewidth]{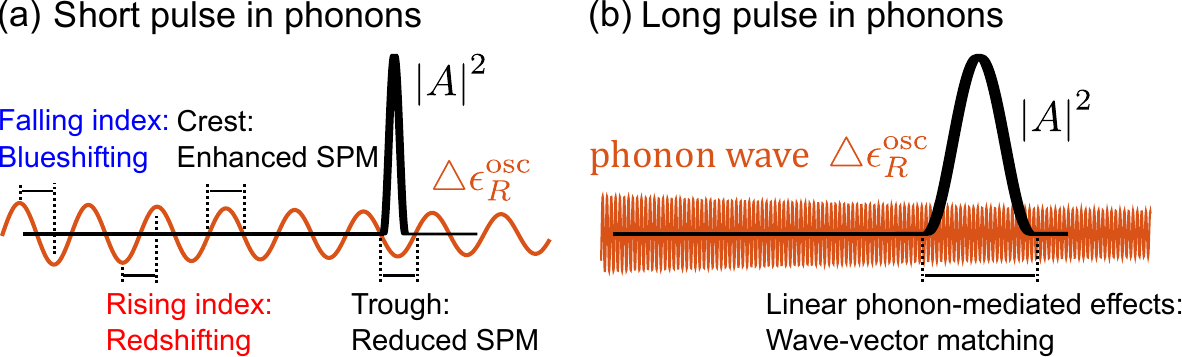}
\caption{Effects of phonons on a pulse. (a) Impulsive phonon effects. (b) Transient and steady-state phonon effects. x-axis represents time. In the steady-state regime with a short phonon lifetime compared to the pulse duration, the pump pulse that generates the phonons must temporally overlap with the probe pulse (Fig.~\ref{fig:overlapping_suppression_ss}).}
\label{fig:Phonon_shortlong_pulse}
\end{figure}

Distinct nonlinear Raman phenomena emerge when an optical field propagates in a medium that contains a pre-existing population of phonons (Fig.~\ref{fig:Phonon_shortlong_pulse}). In the impulsive regime with an ultrashort driving field, the interaction reduces to the familiar femtosecond pump-probe configuration: depending on the relative temporal delay, the pulse can be continuously redshifted, blueshifted, or temporally compressed \cite{Yan1985,Ruhman1987,Dhar1994,Zaitsu2004,Belli2018,Mo2022}. In contrast, the dynamics in the long-pulse transient and steady-state regimes are governed by wave-vector matching between the existing phonons and the discrete Raman interaction \cite{Bauerschmidt2015a,Chen2023,Chen2024}. Because newly-generated (absorbed) phonons are inherently phase-matched to the Stokes (anti-Stokes) process that creates (annihilates) them, this condition can equivalently be interpreted as wave-vector matching between the existing phonons and the phonons to be generated (or absorbed). Such wave-vector matching plays a crucial role in multi-field Raman interactions, for example in two-pulse schemes that employ different wavelengths \cite{Chen2023}, or in cascaded Stokes processes involving different orders that generate phonons with different wave vectors. This process is governed by (Fig.~\ref{fig:schematic_phonon_influcence}; see Supplementary Sec.~10 in \cite{Chen2024} for derivation details):
\begin{subequations}
\begin{align}
\partial_zA^S(z,t) & =-\frac{\omega^S\kappa^SC^{\text{ph}}}{2}\Exp^{-\gamma_2\triangle t}A^P\Exp^{-i\left[\left(\beta^{\text{ph}}-\triangle\beta^S\right)z+\phi^{\text{ph}}\right]} \label{eq:phonon_governing_eqn_S} \\
\partial_zA^{AS}(z,t) & =\frac{\omega^{AS}\kappa^{AS}C^{\text{ph}}}{2}\Exp^{-\gamma_2\triangle t}A^P\Exp^{i\left[\left(\beta^{\text{ph}}+\triangle\beta^{AS}\right)z+\phi^{\text{ph}}\right]}, \label{eq:phonon_governing_eqn_AS}
\end{align} \label{eq:phonon_wave_A2}
\end{subequations}
where $\omega^j$ is the angular frequency of the $j$ [Stokes (S) or anti-Stokes (AS)] wave, $\kappa^j=1/\left(\epsilon_0^2\left[n_{\text{eff}}(\omega^j)\right]^2c^2A_{\text{eff}}(\omega^j)\right)$, $n_{\text{eff}}(\omega)$ is the effective refractive index of the propagating mode, $A_{\text{eff}}(\omega)$ is the effective mode area, $\triangle\beta^j=\beta^P-\beta^j$ is the difference of the propagation constants of the $j$ wave and the pump, and the phonon wave follows ${C^{\text{ph}}(z)\Exp^{-\gamma_2\triangle t}\sin\left(\beta^{\text{ph}}z-\omega_Rt+\phi^{\text{ph}}(z)\right)}$. Unlike the nonlinear Raman amplification governed by the Raman gain, this phonon-mediated Raman generation is a linear process: it is linearly proportional to the pump field $A^P$ and the amplitude of the phonon wave $C^{\text{ph}}$. Because of its linear nature, its generation does not rely on any existing signals to amplify, but directly transfers energy from pump to Raman fields. Therefore, it is an efficient and thresholdless Raman generation mechanism during spontaneous Raman processes \cite{Mridha2019}, which is particularly useful in seeding subsequent nonlinear Raman amplification. In addition, the linear proportionality to complex-valued $A^P$ shows that the generated Raman fields not only are coherent but also inherit the phase structure of the pump pulse. If the wave vector of the seed phonons matches that of the generated phonons, $\beta^{\text{ph}}=\big({\triangle\beta^S}\equiv\beta^{\text{ph,new}}=\beta^P-\beta^S\big)$ [Eq.~(\ref{eq:phonon_governing_eqn_S})], seed phonons can be efficiently amplified, yielding Stokes generation. On the other hand, if the wave vector of the seed phonons can be absorbed while satisfying the conservation of wave vector, $\beta^{\text{ph}}+\triangle\beta^{AS}=0$ and thus $\beta^P+\beta^{\text{ph}}=\beta^{AS}$ [Eq.~(\ref{eq:phonon_governing_eqn_AS})], absorption occurs most efficiently, yielding anti-Stokes generation.

\begin{figure}[!ht]
\centering
\includegraphics[width=\linewidth]{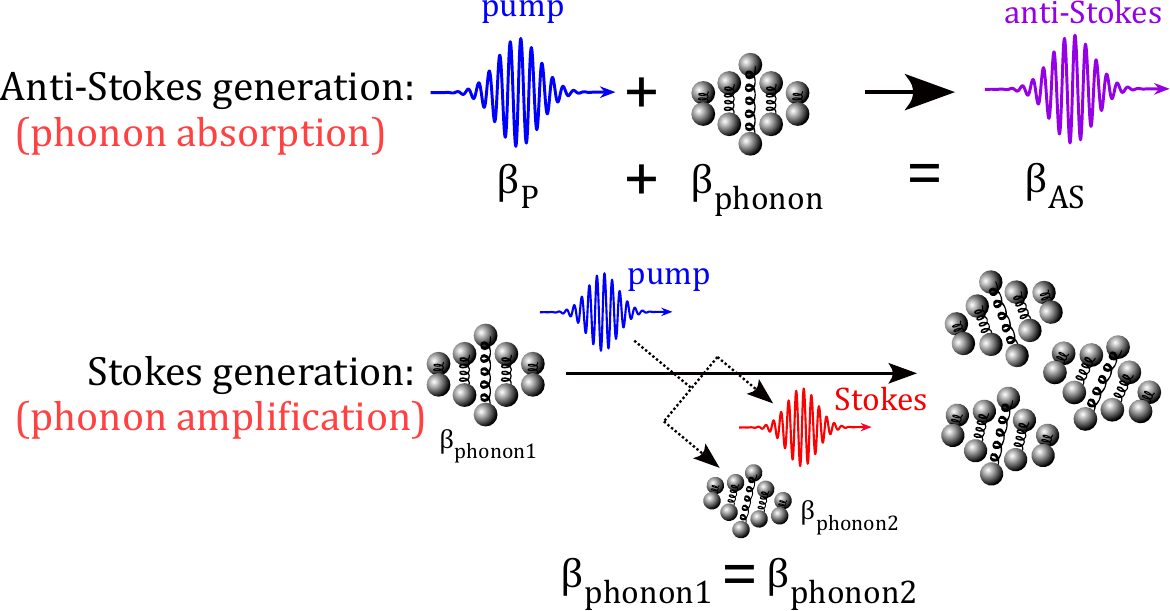}
\caption{Linear phonon-mediated effects on the optical field.}
\label{fig:schematic_phonon_influcence}
\end{figure}

This linear phonon-mediated process provides a complementary description of light-phonon interactions in terms of wave vector or momentum, supplementing the conventional energy-based picture given by $\nu^S+\nu_R=\nu^P$ and $\nu^P+\nu_R=\nu^{AS}$. Phonon absorption appears straightforward, as the associated wave-vector conservation naturally corresponds to the energy relation $\nu^P+\nu_R=\nu^{AS}$. However, a physically-intuitive description of phonon amplification is most naturally formulated in terms of the phonon dynamics themselves, wherein amplification occurs most efficiently when the generated phonons closely resemble the seed phonons.

This process has long been considered as a four-wave-mixing (FWM) process, as in CSRS \cite{Nandakumar2009,Freudiger2008,Ozeki2009,Hamer2024} and CARS \cite{Maker1965,Levenson1972,Duncan1982,Zumbusch1999}. In the traditional steady-state treatment, Raman temporal dynamics can be ignored and the Raman-induced index modulation follows the pulse's temporal profile, so it resembles the instantaneous electronic nonlinearity. In essence, the linear phonon-mediated process does involve four fields, two fields that produce the phonons and two fields in the subsequent phonon amplification or absorption process. Recent investigations in \ce{H2} have demonstrated the importance of non-negligible temporal phonon dynamics due to the long-lived phonons \cite{Tyumenev2022,Aghababaei2023}. In particular, the probe pulse in CARS cannot be too strong; otherwise, its self-induced nonlinear Stokes generation can dominate over the linear anti-Stokes generation [as shown in Fig.~10(d) in \cite{Chen2024}]. This behavior is difficult to reconcile with a description based solely on the nonlinear FWM that also exponentially amplifies the field \cite{Agrawal2013}. As a result, instantaneous FWM is an inaccurate description of this linear phonon-mediated process, and therefore also of the CSRS and CARS.

In the nonlinear-amplification picture, degenerate CARS should not occur, due to Raman gain suppression. When the wave-vector-matching condition ${\triangle\beta_P}={2\beta^P}-\beta^S-\beta^{AS}=0$ is satisfied through a delicate balance of phonon annihilation and creation, the nonlinear Raman gain is zero \cite{Bloembergen1964,Hsieh1974,Perry1985,Duncan1986,Boyd2008}. This originates in the coupling of instantaneous nonlinear phase modulation and Raman gain, and therefore cannot be neglected even in media with strong Raman nonlinearity, where the instantaneous contribution is dominated by the pulse-following Raman-induced index in the long-pulse regime \cite{Chen2024}. In contrast, for linear phonon-mediated interactions, the same condition ${\triangle\beta_P}=0$ corresponds to simultaneous generation of both Stokes and anti-Stokes fields [Eq.~(\ref{eq:phonon_wave_A2})]. Pre-existing phonons, generated by a strong pump and its Stokes field, seed and sustain the scattering processes, overcoming the nonlinear Raman gain suppression. This is precisely the mechanism that enables degenerate CARS under perfect wave-vector matching.

Even considering the new perspective offered by the linear phonon-mediated processes in CARS and CSRS, we emphasize that the traditional frequency-domain, FWM-based explanation provides an intuitive understanding if the Raman process is in the steady-state regime \cite{Rigneault2018}. The traditional interpretation and mathematical formulation are valid because (1) they assume instantaneous steady-state Raman operations, where the phonon temporal dynamics can be reasonably ignored, and (2) the weak-probe condition in CARS/CSRS resembles the weak-pump assumption [$A^P$ in Eq.~(\ref{eq:phonon_wave_A2})] in the analysis of the linear phonon-mediated process where the instantaneous nonlinearity from both the electronic and Raman effects is ignored \cite{Chen2024}. That is, the traditional formulation happens to coincide with the steady-state linear phonon-mediated process. The two approaches yield the same intensity dependence for the Raman fields \cite{Rigneault2018}: with $C^{\text{ph}}\sim{\triangle\epsilon_R^{\text{osc}}}\sim{A^{P1}\left(A^{S1}\right)^*}$, the anti-Stokes intensity in CARS $I^{AS}\propto{I^{P1}I^{S1}I^{P}}$ under conditions of \emph{incoherent} growth [Eq.~(\ref{eq:phonon_governing_eqn_AS})], with $I^j=\abs{A^j}^2$, and it is proportional to $\left(I^P\right)^2I^{S1}$ for degenerate CARS where ${P1}=P$. On the other hand, CARS and CSRS require the linear phonon-mediated processes to obtain a physically-intuitive time-domain description of their dynamics outside the steady-state regime. Although a frequency-domain description is also possible by using the convolution theorem, $\mathfrak{F}\left[\triangle\epsilon_R(z,t)A(z,t)\right]\propto\mathfrak{F}\left[\triangle\epsilon_R(z,t)\right]\ast\mathfrak{F}\left[\triangle A(z,t)\right]$, its applicability is essentially limited to Stokes/anti-Stokes (weak) signal detection, usually with a short interaction length \cite{Bartels2021}. With short propagation, where cumulative phase effects are negligible, dispersion is usually considered negligible in prior analyses. However, once dispersion is ignored, the underlying wave-vector-matching condition is lost, and the constructive or destructive cumulative growth dynamics of the Raman-generated fields cannot be captured [Eq.~(\ref{eq:phonon_wave_A2})]. In fact, dispersion can never be ignored; it determines the Raman gain that underlies Raman growth at every point along the propagation \cite{Bloembergen1964,Chen2024}, rather than merely contributing a cumulative phase effect. Only in the nonphysical limit $\abs{\triangle\beta_P}\rightarrow\infty$ (instead of negligible dispersion ${\triangle\beta_P}=0$) does the Raman gain become independent of wave-vector matching, \ie far from the condition for Raman gain suppression [see Eq.~(S203) in \cite{Chen2024}], so that it depends solely on $\mathfrak{F}\left[\triangle\epsilon_R(z,t)\right]$. Consequently, a frequency-domain formulation that retains only the nonlinear terms is valid exclusively under operational conditions where phonons drive the Raman response of a weak probe field -- conditions that correspond fundamentally to the linear phonon-mediated process. Since Raman evolution intrinsically comprises both the linear phonon-mediated response and the nonlinear Raman-gain dynamics, removing the latter eliminates the wave-vector dependence during propagation and thereby justifies analyses that consider only the nonlinear terms.

Similar to CSRS and CARS, molecular (phase) modulation is not a distinct mechanism, but another manifestation of the linear phonon-mediated process. This is the process by which an incident field is modulated by previously-excited coherent molecular motions, namely phonons. This approach has been employed for the generation of few-cycle pulses and frequency combs \cite{Harris1998,Sokolov2002,Zhavoronkov2002,Bustard2008,Baker2011}, or for frequency upconversion \cite{Bauerschmidt2015a,Mridha2016,Mori2018,Tyumenev2022}. The former arises from thresholdless coherent generation of multiple Raman orders, and the latter follows directly from Eq.~(\ref{eq:phonon_governing_eqn_AS}). Eq.~(\ref{eq:phonon_wave_A2}) is the physically-intuitive counterpart to the recent quantum description of molecular modulation in \cite{GonzalezRaya2025}; however, the quantum treatment relies on the steady-state assumption, which neglects phonon temporal dynamics. In contrast, Eqs.~(\ref{eq:phonon_wave_A2}) remain valid across timescales, including the case of nonlocal interactions between two temporally-separated pulses, which we will discuss in the next section. Quantum state preservation and entanglement arise from the linear proportionality to both the phonon wave and the converting pump field [Eq.~(\ref{eq:phonon_wave_A2})], which enforces coherent correlation among all participating fields \cite{Tyumenev2022,Aghababaei2023,GonzalezRaya2025}.

\begin{figure}[!ht]
\centering
\includegraphics[width=\linewidth]{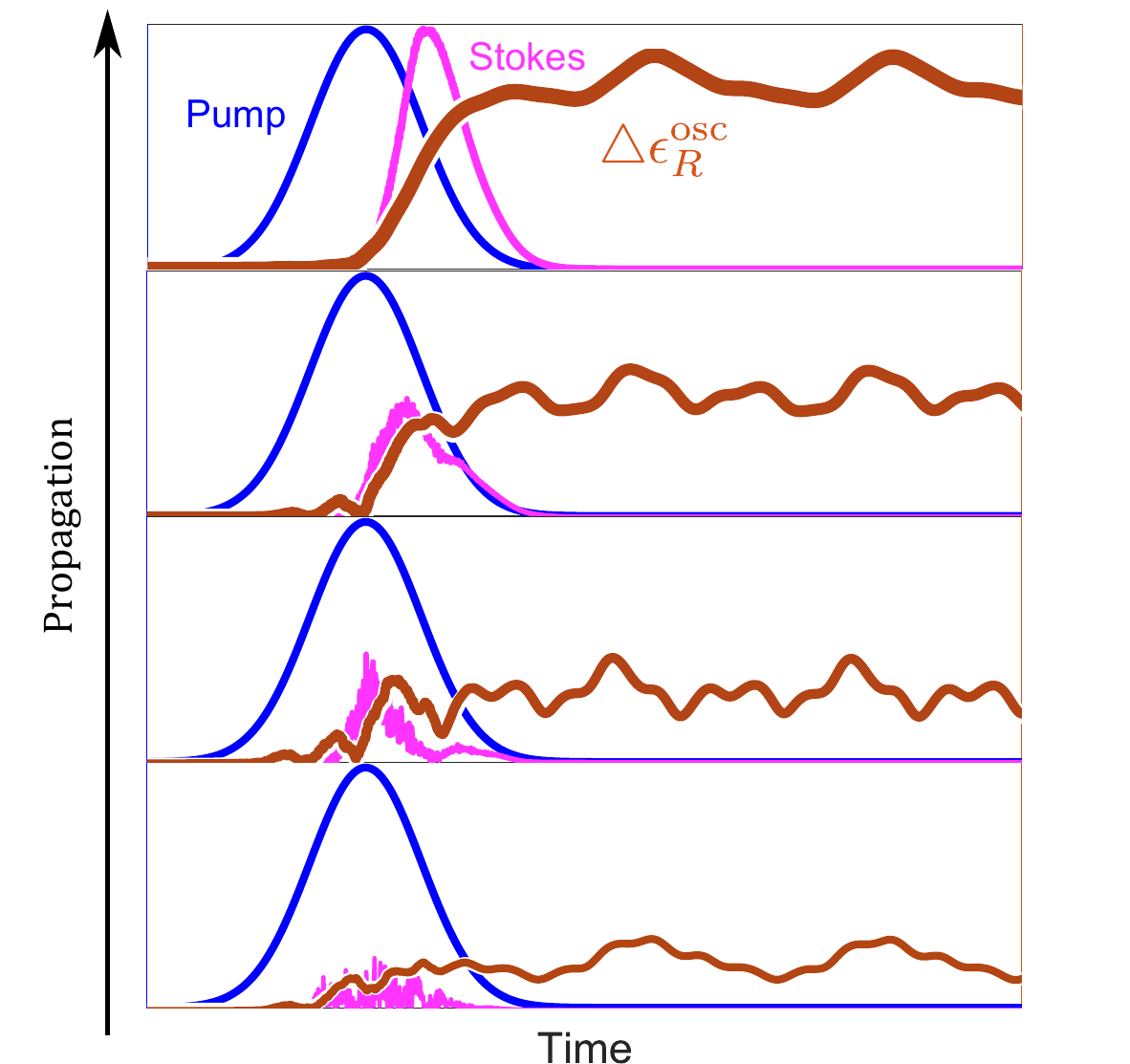}
\caption{Build-up of Stokes field (pink) from spontaneous Raman generation in the transient regime, with the intensity of each field normalized for better visualization. The oscillatory part of the Raman-induced index modulation ($\triangle\epsilon_R^{\text{osc}}$, orange), \ie the phonon wave, is isolated by filtering out the zero-frequency component. The Stokes field is the accumulation of Stokes increments ${i\triangle\epsilon_R^{\text{osc}}(z,t)A(z,t)}$ over the propagation distance. The Stokes intensity grows with propagation, with the new coherent part overriding the initial incoherent spontaneous part. The simulation assumes vibrational Raman generation in \ce{H2}, which has an \SIadj{8}{\fs} period, so, for visualization the envelope/amplitude of the phonon wave is plotted. The input pulse duration is \SI{3}{\ps}, much shorter than the nanosecond dephasing time, so the interaction is in the transient regime.}
\label{fig:transient_spontaneous}
\end{figure}

The linear phonon-mediated processes also fill the existing gap in understanding of coherent Stokes generation from spontaneous Raman scattering in the transient regime (Fig.~\ref{fig:transient_spontaneous}) \cite{Carman1970}. Unlike incoherent Stokes fields that grow from steady-state spontaneous Raman generation, transient spontaneous Raman generation produces coherent Stokes fields that exhibit the same temporal phase structure as the pump pulse. To investigate this, we performed simulations with a unidirectional pulse propagation equation (UPPE) that includes realistic Raman response functions for gases \cite{Chen2024}, and the results are shown in Fig.~\ref{fig:transient_spontaneous}. Initially, the Stokes field is seeded by vacuum fluctuations \cite{Milonni1994}. Due to the incoherence of vacuum fluctuations, the beating with the pump field fluctuates, which leads to an incoherent phonon response. The Stokes increment $\triangle\epsilon_R^{\text{osc}}(z,t)A(z,t)$ during propagation \cite{Chen2026} is therefore dominated by the pump field (bottom panel in Fig.~\ref{fig:transient_spontaneous}). However, the finite phonon response time (due to the Raman period) results in a smoothed phonon wave with increased coherence, which then drives more coherent Stokes generation through the linear phonon-mediated process. This new coherent Stokes wave overrides the initial incoherent Stokes wave, and seeds the subsequent nonlinear amplification, eventually producing a coherent Stokes field that inherits the temporal phase from the pump [Eq.~(\ref{eq:phonon_governing_eqn_S})]. Moreover, the growing phonon wave leads to a Stokes increment that is delayed to the trailing edge of the pump pulse. Traditional frameworks based solely on nonlinear Raman amplification, in which the incoherent Stokes background is amplified, fail to account for the emergence of coherence in this process. They omit the phonon temporal dynamics that are crucial for coherence generation, and are thus applicable only in the steady-state regime. \etal{Carman} analyzed transient Stokes generation by assuming \textit{a priori} that the Stokes and pump fields share the same phase structure as they evolve in a high-gain environment \cite{Carman1970}. At first glance, this explanation appears physically reasonable: high-gain Raman amplification should preferentially reinforce components capable of canceling $\left(A^P\right)^*A^S$ in the Raman-gain integral [see Eq.~(15) in \cite{Chen2024}]. However, from a mathematical standpoint, if $A^S(z=0,t)$ originates from noise, the noisy integral does not clearly demonstrate that $A^S(z,t)$ will evolve into a coherent field, let alone one that specifically inherits the pump phase. Only with the introduction of the linear phonon-mediated dynamics is this coherence growth clearly explained.

Eqs.~(\ref{eq:phonon_wave_A2}) isolate the linear phonon-mediated contribution from the full Raman response $\left[\partial_zA(z,t)\right]_{\mathcal{R}}\propto {i\left[\triangle\epsilon_R(z,t)A(z,t)\right]}$, which clarifies the specific role this pathway plays in Raman temporal dynamics. Because the phonon wave is continuously driven by the optical fields, its amplitude $C^{\text{ph}}$ and phase $\phi^{\text{ph}}$ vary with propagation distance, whereas the governing phonon wave vector $\beta^{\text{ph}}$ remains fixed, determined by the dispersion relation in the medium. Therefore, Eqs.~(\ref{eq:phonon_wave_A2}) are not independent evolution equation, but coupled equations that need to be solved in conjunction with with the phonon dynamics equation for ${C^{\text{ph}}(z)}$ and ${\phi^{\text{ph}}(z)}$. The complete dynamics can also be obtained from a single equation (\eg as in the UPPE \cite{Chen2024}) with the full Raman formulation ${\triangle\epsilon_R(z,t)A(z,t)}$ where $A(z,t)$ explicitly contains both the field components that excite the phonons and those that are to be converted. The newly-introduced equations thus complement, rather than supplant, the conventional nonlinear description by highlighting the linear phonon-mediated pathway to make its physical role more transparent.

\section{Phonon-controlled Raman scattering}
The linear phonon-mediated process underlies early cascaded Raman generation. A Stokes field can act as a new pump field and interact with existing phonons generated by the first-order Stokes scattering, which drives thresholdless Raman generation of various orders through linear phonon-mediated processes. Simultaneous growth of multiple Raman sidebands reduces the energy directed into a single order. Continuous bidirectional nonlinear coupling among multiple Raman signals degrades the temporal fidelity of each individual signal. Existing techniques for suppression of the Raman cascade involve controlling Raman gain suppression by phase-matching Stokes and anti-Stokes fields \cite{Bloembergen1964}, adding loss to the Raman band \cite{Fini2006}, inhibiting Raman feedback in a cavity \cite{Xu2017}, or arranging for competition with modulation instability \cite{Couderc2005}; however, they completely suppress the Raman process. How to cleanly suppress generation of only the higher orders, or even target a specific order, remains unknown.

Suppression of phonon-mediated Raman generation can mitigate the Raman cascade when it is undesirable. Phonon-mediated processes depend on the wave-vector matching between the phonon wave vector and the corresponding Raman generation. When the wave-vector mismatch is nonzero [Eqs.~(\ref{eq:phonon_wave_A2})], destructive interference occurs over extended propagation distances. Stronger destructive interference can also be achieved by directly increasing the wave-vector mismatch, which may be implemented by adjusting the gas pressure, introducing a buffer gas, or tailoring the hollow-core fiber structure. We limit the discussion to the second-order Stokes generation as the start of a cascaded process. To drive the second-order Stokes process with phonons generated from the first-order process, the phonon wave vector $\beta^{\text{ph},S_1}$ should match that from the second-order, $\beta^{\text{ph},S_1}=\left(\beta^{S_1}-\beta^{S_2}\equiv\beta^{\text{ph},S_2}\right)$ [Eq.~(\ref{eq:phonon_governing_eqn_S})], where $S_j$ represents the $j$th-order Stokes field. Because $\beta^{\text{ph},S_1}\equiv\beta^P-\beta^{S_1}$, this cascaded wave-vector-matching relation is equivalent to the conventional wave-vector-matching relation but treating the first-order Stokes wavelength as a new pump:
\begin{equation}
\triangle\beta_{S_1}=2\beta^{S_1}-\beta^P-\beta^{S_2}=0.
\label{eq:beta_cascaded}
\end{equation}

\begin{figure}[!ht]
\centering
\includegraphics[width=\linewidth]{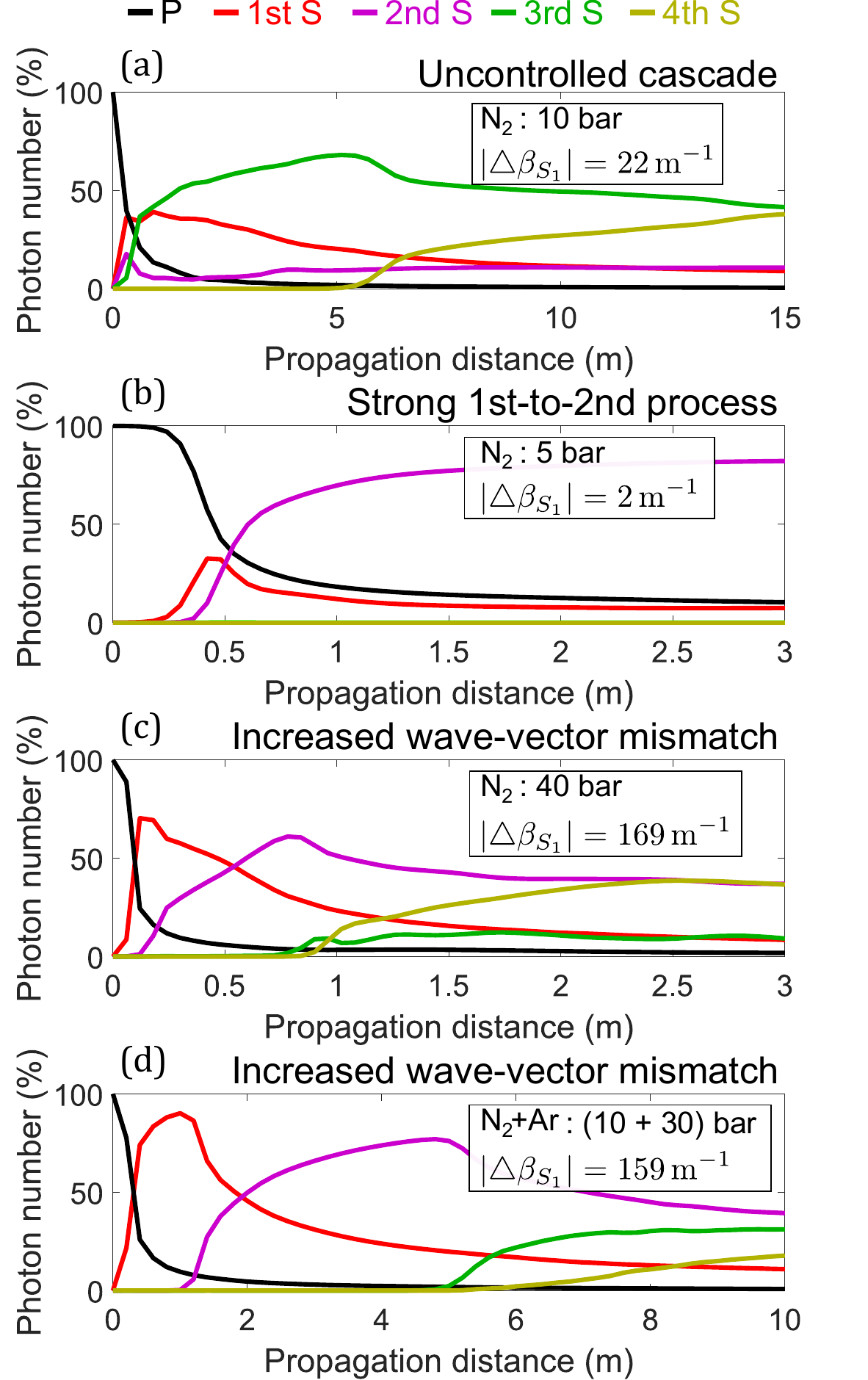}
\caption{Demonstration of phonon control and suppression of early-stage cascaded Raman generation with numerical simulations of seeded vibrational Stokes generation in \ce{N2}-filled hollow-core fiber. Fiber and pulse parameters follow \cite{Gao2022}, but the pump pulse energy is adjusted to \SI{4}{\micro\joule}. A weak \SIadj{1}{\nano\joule} and \SIadj{20}{\ps} seed at the first Stokes wavelength propagates with a temporal offset of \SI{10}{\ps} toward the leading edge of the \SIadj{20}{\ps} pump pulse. (a) Early cascaded Raman generation. (b) Enhanced energy transfer from the first to the second Stokes order. (c,d) Suppression of early-stage cascaded Raman generation through increasing wave-vector mismatch. At the photon-number maximums of the first and second Stokes orders, the energies residing in all other orders remain negligible. Despite suppression of early cascaded Raman growth for these two orders, the third and fourth orders grow simultaneously due to small $\abs{\triangle\beta_{S_3}}$. ${\triangle\beta_{S_1}}$ represents the wave-vector mismatch for phonons generated by the first- and the second-order Stokes processes [Eq.~(\ref{eq:beta_cascaded})].}
\label{fig:overlapping_suppression}
\end{figure}

To verify this approach, we simulated vibrational Raman scattering in a \ce{N2}-filled hollow-core fiber with the aforementioned UPPE. The fiber and pulse parameters match those of an experiment that generates a Raman frequency comb \cite{Gao2022}. A linearly-polarized input pulse was accompanied by a weak Stokes pulse. Despite seeding the first Stokes field, significant early-stage and thresholdless cascaded vibrational Raman scattering occurs due to the small wave-vector mismatches with first-order phonons across multiple Stokes orders [small $\abs{\triangle\beta_{S_1}}$ and $\abs{\triangle\beta_{S_2}}$ in Fig.~\ref{fig:overlapping_suppression}(a)]. Stokes pulses not only have reduced energy but also suffer degraded temporal quality, manifested as pronounced temporal substructures resulting from energy exchange among the various Stokes orders (illustrated in Supplementary Sec.~S1). By adjusting the gas pressure to \SI{5}{\bar} so that $\abs{\triangle\beta_{S_1}}\approx0$, efficient energy transfer from the first to the second Stokes order is enabled [Fig.~\ref{fig:overlapping_suppression}(b)]. In this case, the increased wave-vector mismatch for the phonon-mediated second- to third-order process (\SI{55}{\m^{-1}}) causes Raman generation to terminate at the second Stokes order, and results in highly-efficient conversion to that order.

Complete suppression of phonon-mediated processes through an increased wave-vector mismatch over multiple orders is illustrated in Figs.~\ref{fig:overlapping_suppression}(c) and (d) [with only the first- to second-order ${\triangle\beta_{S_1}}$ displayed; Eq.~(\ref{eq:beta_cascaded})]. The enhanced wave-vector mismatch yields strong destructive interference, thereby inhibiting the phonon-mediated Raman generation and preventing the simultaneous growth of multiple Stokes orders. Under these conditions, the Raman process becomes effectively sequential: the Stokes signal of a given order grows only after the preceding order has reached its maximum, and this pattern repeats for each successive order. A Raman pulse that simultaneously achieves high energy and high temporal fidelity can thus be generated.

\begin{figure}[!ht]
\centering
\includegraphics[width=\linewidth]{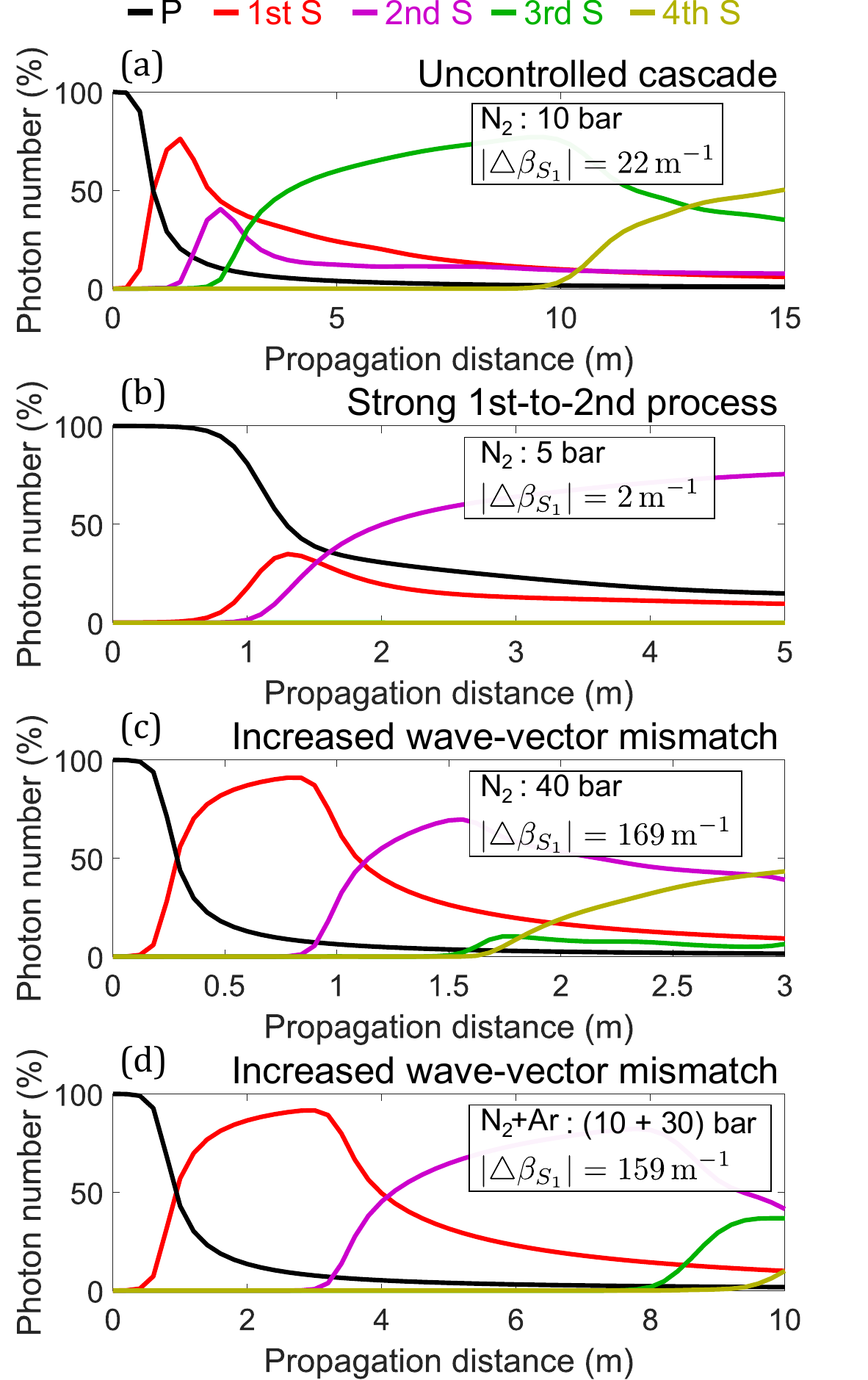}
\caption{Demonstration of phonon control and suppression of early-stage cascaded Raman generation in the steady-state versions of processes in Fig.~\ref{fig:overlapping_suppression}. The Raman dephasing times were artificially reduced by \num{100} times to reach the steady-state regime, with a \num{6}-fold increase in the vibrational Raman strength to compensate the reduced Raman amplification (which occurs after the linear phonon-mediated seeding process). It is important to increase the Raman strength here, rather than increase the propagation distances, because extended propagation (under a weaker nonlinear process) enhances the phase mismatch in the linear phonon-mediated processes [Eq.~(\ref{eq:phonon_wave_A2})], which would be another route for suppressing early cascaded Raman generation. The Stokes seed pulse was adjusted to \SI{30}{\ps} duration, with no temporal offset from the pump, to ensure efficient steady-state Stokes seeding.}
\label{fig:overlapping_suppression_ss}
\end{figure}

To explore control of phonon-mediated processes in the steady-state regime, we artificially reduced the Raman dephasing times, while keeping all other parameters the same as in the examples of Fig.~\ref{fig:overlapping_suppression}. The resulting dynamics are essentially unchanged (Fig.~\ref{fig:overlapping_suppression_ss}). This demonstrates that even in the steady-state limit, where the phonon temporal dynamics might seem to be negligible, the evolution is intrinsically transient: each temporal slice of the field is influenced by phonons generated by earlier slices. More broadly, the action of the linear phonon-mediated process across all temporal regimes underscores that, irrespective of operating conditions, Raman scattering is fundamentally an ultrafast phonon-driven phenomenon governed by femtosecond phonon dynamics.

As another example of the benefits of controlling the phonon-mediated processes, we simulated the two-pulse approach to Raman-shifting, which is employed to suppress Raman spectral narrowing (see Supplementary Fig.~S1) \cite{Konyashchenko2007,Konyashchenko2008,Didenko2015,Vicario2016,Konyashchenko2017}. When a chirped input pulse undergoes transient Raman scattering, the Raman gain is strongest near its trailing edge \cite{Chen2024}, so only the trailing portion of the pulse is efficiently converted. This yields a Stokes pulse with a bandwidth narrower than that of the pump. In the two-pulse configuration, however, the phonon wave generated by the first pulse can linearly drive Raman generation across the entire duration, and thus the spectrum, of the chirped second pulse [Eqs.~(\ref{eq:phonon_wave_A2})]. In the conditions considered here (under \SIadj{10}{\bar} \ce{N2}), phonons generated by the first pulse exhibit only small wave-vector mismatch with the second-order Stokes process driven by the second pulse. As a result, both the first- and second-order Stokes components grow simultaneously, giving rise to undesired early cascaded Raman generation. An increased wave-vector mismatch suppresses the cascaded process similarly. Most importantly, it shows that the two-pulse approach does not consistently yield a good first-order Stokes pulse. The effectiveness of this approach is tightly constrained by the wave-vector-matching conditions between phonons and subsequent Raman processes, which ultimately determine whether a phonon-mediated process is enhanced or suppressed.

\section{Comparison with prior Raman works}\label{sec:Comparison}
Since its discovery in 1923 \cite{Smekal1923} and its experimental confirmation in 1928 \cite{Raman1928,Raman1928b,Raman1928a,Landsherg1928,Landsherg1928a,Landsherg1928b}, Raman scattering has been the subject of extensive investigation. Prior studies have also examined Raman-induced index modulation to explain the Raman dynamics, including CSRS/CARS. The purpose of this section is to articulate the novel features that distinguish the present work from the substantial body of existing research.

Given the widespread use of four-photon CSRS/CARS, we first outline the aspects of the present work that extend beyond the conventional CSRS/CARS framework, and subsequently return to discuss their implications for CSRS/CARS in detail.
\begin{enumerate}
\item \textbf{Raman phonons:} This paper, for the first time, discusses the role of Raman phonons in the index change. Raman phonons have usually been considered a quantum-mechanical by-product of the optical Stokes process. Only with deep understanding can the underlying physics be uncovered, such as the linear phonon-mediated process [Eq.~(\ref{eq:phonon_wave_A2}) and Fig.~\ref{fig:schematic_phonon_influcence}] that underlies the temporal phase-copying effect and coherent pulse generation in spontaneous Raman generation in the transient Raman regime (Fig.~\ref{fig:transient_spontaneous}).
\item \textbf{Application to rotational SRS:} Sec.~\ref{sec:Raman_phonons} shows that the formulation based on the index change applies equally well to both vibrational and rotational Raman scattering. Traditional descriptions mostly focus on vibrational Raman scattering \cite{Korn1998,Nazarkin1998,Nazarkin1999,Boyd2008}; 
when rotational Raman scattering is considered, complicated quantum rotational dynamics are employed \cite{Martin1988,Nibbering1997,Chen2007,Hoque2011,Langevin2019} and the physics is not connected to the index change. Thus, the underlying physics is obscured.
\item \textbf{Full time-domain framework:} Traditionally, Raman scattering has been analyzed within a frequency-domain framework, even in the context of index modulation \cite{Bartels2021}, and this is especially common in CSRS/CARS research. However, Raman scattering is inherently a spectrotemporal process whose behavior depends sensitively on the relevant timescales. A full time-domain formulation provides a more physically-intuitive description of the underlying mechanism, such as the optical Raman-phonon effect in Fig.~\ref{fig:Phonon_shortlong_pulse}, as well as Raman-induced temporal distortion \cite{Chen2026} and nonlinear index smoothing \cite{Wang2026a} arising from the inertial nature of the response. In the time domain, Raman scattering fundamentally originates from the optical interaction with the material’s delayed nonlinear response, which induces a real-valued index variation $\triangle\epsilon_R(z,t)$. Only after Fourier transformation does this response become complex-valued in the frequency domain. The need to treat its real and imaginary parts separately (corresponding to phase and amplitude modulations, respectively) introduces additional analytical complexity.
\item \textbf{Temporal regimes:} The discussion here, in combination with prior works \cite{Chen2024,Chen2026}, comprehensively covers Raman dynamics across different temporal Raman regimes. For example, in long-pulse regimes, unlike traditional approaches that consider only the Raman oscillatory part $\Exp^{i\left(\beta^{\text{ph}}z-\omega_Rt\right)}$ for the amplitude modulation, the pulse-following part is included, which is crucial in Raman-enhanced SPM \cite{Belli2015,Hosseini2018,Konyashchenko2019,Beetar2020}. The optical interaction with Raman phonons depends on rising or falling index in short-pulse regimes, or wave-vector matching in long-pulse regimes (Fig.~\ref{fig:Phonon_shortlong_pulse}).
\item \textbf{Highly-efficient single-order Raman generation:} Building on the new understanding of Raman phonons, we, for the first time, propose a technique to suppress early-stage cascaded Raman generation. This enables Raman generation to only a single order, rather than complete suppression of Raman generation in prior works \cite{Bloembergen1964,Fini2006,Xu2017,Couderc2005}.
\item \textbf{Complete Raman picture:} A complete picture of Raman scattering is established (Supplementary Fig.~S2), involving the newly-introduced linear phonon-mediated process [Eq.~(\ref{eq:phonon_wave_A2}) and Fig.~\ref{fig:schematic_phonon_influcence}] and the well-known nonlinear Raman gain dynamics \cite{Bloembergen1964,Chen2024}.
\end{enumerate}

Having outlined the novel features compared to the conventional CSRS/CARS framework, we now detail the application of the linear phonon-mediated process to CSRS/CARS and state the new conclusions:
\begin{enumerate}
\item \textbf{Evolution dynamics:} Most prior CSRS/CARS studies neglect the evolution dynamics that fundamentally govern Raman growth; they consider only the form of each nonlinear term under a chosen launch condition. As discussed earlier, dispersion plays a central role in the Raman gain that continuously shapes the Raman dynamics, rather than merely contributing a cumulative phase that appears negligible over short propagation distances. Once this dispersion dependence is reinstated, the evolution assumed in existing CSRS/CARS treatments no longer holds. For example, CARS should not occur due to the Raman gain suppression under the nonlinear Raman-gain picture \cite{Bloembergen1964,Chen2024}.
\item \textbf{Coincidental conditions:} Raman scattering consists of both the linear phonon-mediated process and the nonlinear Raman-gain dynamics. Because Raman gain is always active during propagation, conventional CSRS/CARS analyses, which neglect this gain, break down. Their validity is recovered only in regimes where the Raman-gain contribution can be neglected and the linear phonon-mediated process dominates. This occurs under conditions of pre-existing or co-existing phonons together with a weak probe, which are precisely the typical operational conditions of CSRS/CARS.
\end{enumerate}

\section{Conclusion}
In conclusion, we have illuminated the role of Raman phonons through the unifying framework of the Raman-induced index modulation $\triangle\epsilon_R(z,t)$, spanning different temporal regimes, different media, and distinct Raman interaction types (see Supplementary Fig.~S2 for a concluding figure). This analysis helps elucidate linear phonon-mediated processes, and provides a complementary wave-vector-based perspective alongside the traditional energy-based description. This framework clarifies the underlying mechanisms of CSRS, CARS, molecular modulation, and coherent spontaneous transient Raman generation. Finally, we introduced a phonon-controlled approach for deliberately enhancing or inhibiting the linear phonon-mediated process, which enables the suppression of early-stage cascaded Raman scattering for highly-efficient Stokes generation at a specific order in numerical simulations. The influence of Raman phonons across distinct time scales underscores the inherently ultrafast time-dependent nature of Raman scattering, an aspect fundamentally missed by the traditional steady-state analysis.

\begin{backmatter}
\bmsection{Funding} National Institutes of Health (R01EB033179, U01NS128660), the Office of Naval Research (N00014-19-1-2592).

\bmsection{Acknowledgments} Y.-H.C. was partially supported by a Mong Fellowship from Cornell Neurotech.

\bmsection{Disclosures} The authors declare no conflicts of interest.

\bmsection{Data availability} The code and the data used in this article have been made publicly available at \url{https://github.com/AaHaHaa/gas_UPPE}.

\bmsection{Supplemental document} See Supplement~1 for supporting content. 

\end{backmatter}

\bibliography{reference}

\bibliographyfullrefs{reference}

\end{document}


\maketitle

\newpage 

\section{Supplementary figures}
\label{sec:Supplementary_figures}

\begin{figure}[!ht]
\centering
\includegraphics[width=0.8\linewidth]{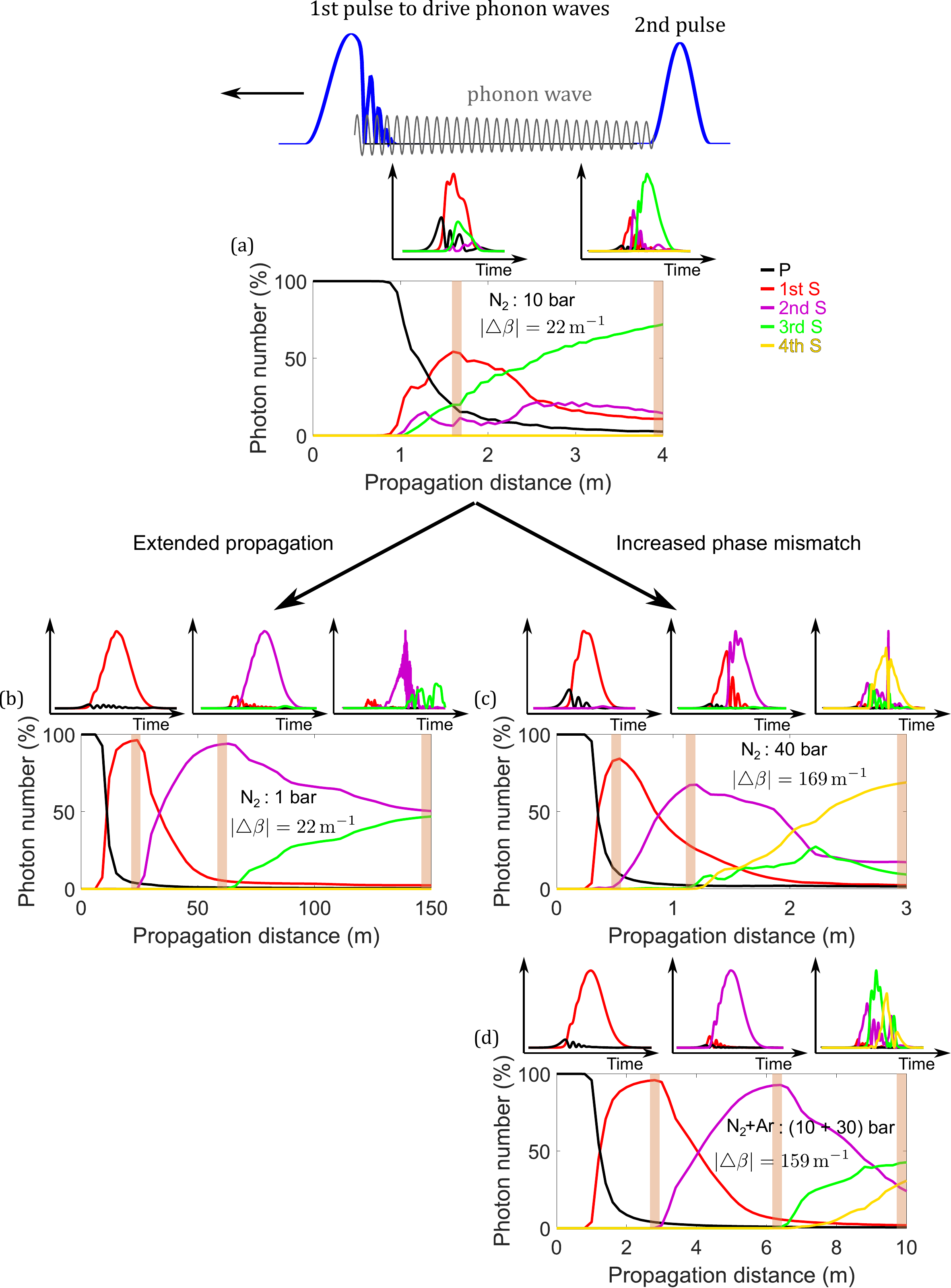}
\caption{Demonstration of phonon control and suppression of early-stage cascaded Raman generation with two-pulse vibrational Raman generation in \ce{N2}-filled hollow-core fiber. (a) Early cascaded Raman generation. Suppression of early-stage cascaded Raman generation through (b) extended propagation or (c,d) increasing wave-vector mismatch. Temporal profiles of different signals (on top of each figure) are shown at the propagation distances corresponding to the colored regions.}
\label{fig:two-pulse_Stokes}
\end{figure}

\clearpage

\begin{sidewaysfigure}[!ht]
\centering
\includegraphics[width=\linewidth]{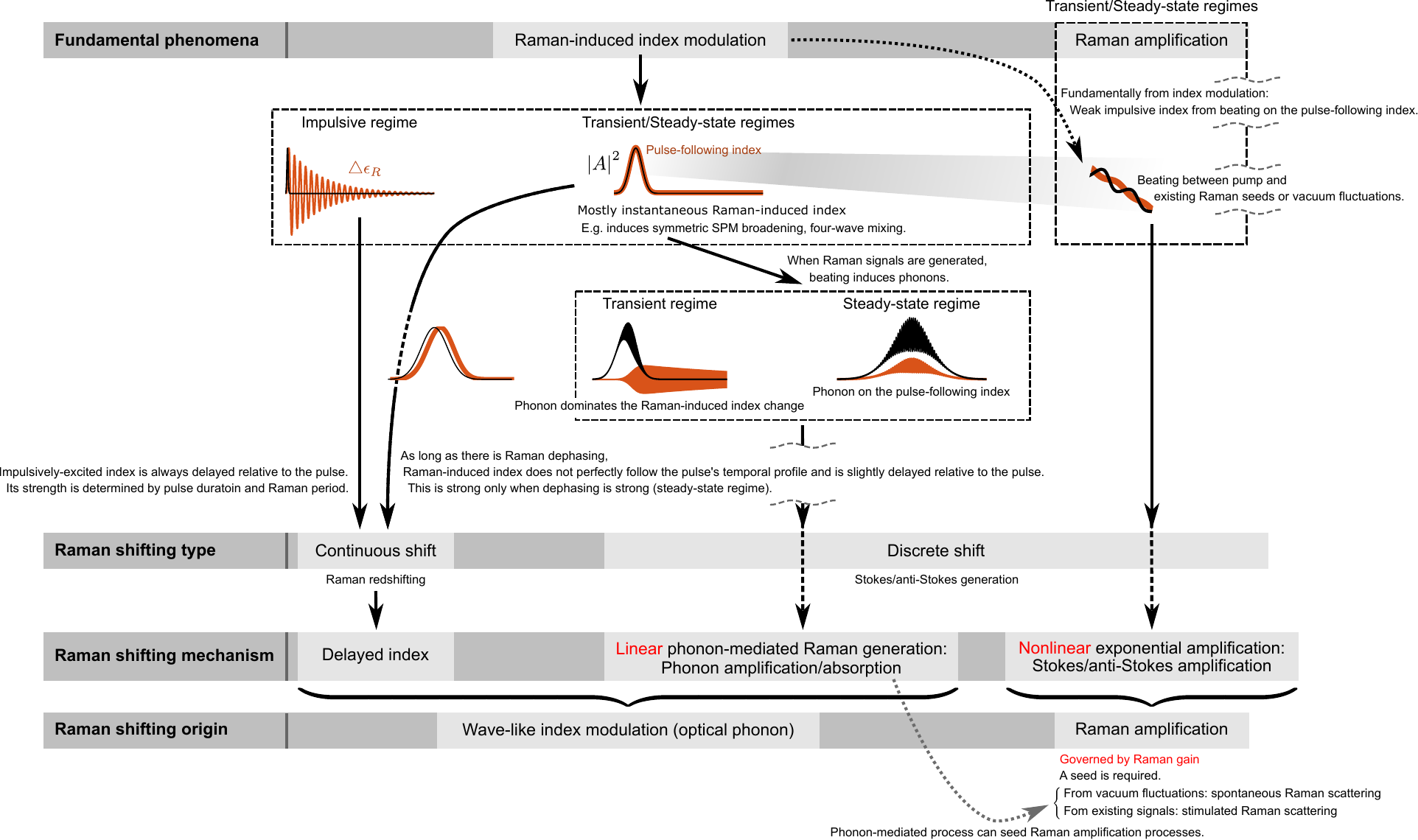}
\caption{Complete picture of Raman scattering. ``Raman shifting'' on the left includes continuous redshifting and discrete Stokes/anti-Stokes generation, whereas ``fundamental phenomena'' includes all Raman-induced effects, involving Raman shifting and Raman-enhanced SPM, etc.}
\label{fig:Raman_regimes_supplement}
\end{sidewaysfigure}

\clearpage
\section{Discrete Raman generation}
\label{sec:Discrete_Raman_generation}
Traditional understanding of Raman phonon lies in long-pulse Raman regimes where frequency shifting manifests as discrete Raman sideband generation, and the Raman-induced index modulation ${\triangle\epsilon_R(t)}$ has clear frequency peaks at Raman transition frequencies $\left(n\omega_R\right)$, where $n\in\mathbb{Z}$, far from its zero-frequency part. Here, we elaborate more on the origin and the resulting nonlinear Raman effects of these oscillatory \emph{phonons}.

\begin{figure}[!ht]
\centering
\includegraphics[width=0.6\linewidth]{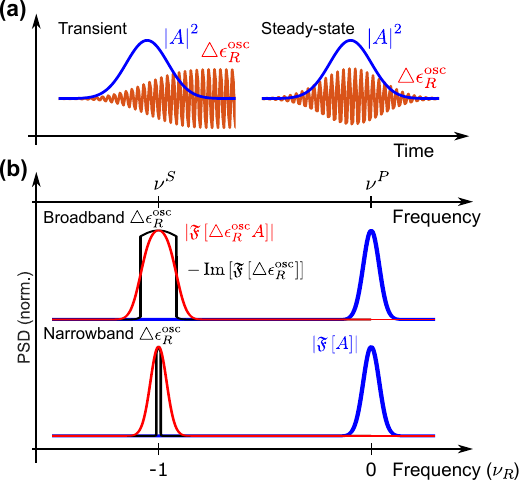}
\caption{Long-pulse spectrotemporal Raman dynamics. (a) Temporal evolution of the beating-induced index modulation $\triangle\epsilon_R^{\text{osc}}(t)$. Index modulation extends beyond the pulse in the transient regime due to $\tau_0\ll T_2$, leading to integrated-energy-dependent and thus stronger Stokes generation in the trailing edge \cite{Chen2024}. (b) Illustration of Raman spectral shaping. If ${\triangle\epsilon_R^{\text{osc}}}$ is narrowband (approximately a Dirac delta function in the frequency domain), the spectral convolution, from temporal multiplication, yields a nonlinear Stokes increment ${\triangle\epsilon_R^{\text{osc}}(t)A(t)}$ that has the same bandwidth as the pump pulse $A(t)$; otherwise, the Stokes spectrum is broadened. In fact due to the temporal convolution in $\triangle\epsilon_R^{\text{osc}}(t)$, the bandwidth of $\triangle\epsilon_R^{\text{osc}}$ is approximately the minimum of the Raman-gain bandwidth $\mathfrak{F}\left[R\right]$ and the bandwidth of the beating amplitude $\mathfrak{F}\left[C^{PS}\right]$, and therefore of the pump and Stokes fields themselves. In other words, the bandwidth of $\triangle\epsilon_R^{\text{osc}}$ (black lines) can only be equal to or narrower than the pump bandwidth (blue line), implying that the maximum bandwidth of $\triangle\epsilon_R^{\text{osc}}(t)A(t)$ is $\sqrt{2}$ times the pump bandwidth. This shows that the initial Stokes growth draws energy from anywhere between ${1/\sqrt{2}}$ and the full temporal extent of the pump pulse. The spectral phase information within ${\mathfrak{F}\left[\triangle\epsilon_R^{\text{osc}}(t)A(t)\right]}$ (red lines) is governed by the Raman temporal dynamics: it represents the Stokes field at the trailing edge of the pump in the transient regime or at the pump's temporal center with the highest instantaneous intensity in the steady-state regime [also see (a)] \cite{Chen2024}. $\nu^{P/S}$ is pump (P) or Stokes (S) frequency.}
\label{fig:long-pulse}
\end{figure}

Discrete Raman sidebands originate from vacuum fluctuations or temporally-overlapped Raman seeds whose beating drives index oscillations at the Raman transition frequency
\begin{align}
{\triangle\epsilon_R(t)} & \propto{R(t)\ast\abs{A^P(t)+A^S(t)\Exp^{i\triangle\omega t}}^2} \nonumber \\
& \approx{R(t)\ast\left[\abs{A^P(t)}^2+\abs{A^S(t)}^2+C^{PS}(t)\sin\left(-\triangle\omega t\right)\right]},
\end{align}
where $C^{PS}(t)\sim\mathfrak{F}\left[A^P(t)\left(A^S(t)\right)^*\right]$ denotes the beating amplitude, and $A^{P/S}$ is the pump (P) or Stokes (S) field with ${\triangle\omega}=\omega^S-\omega^P={-\omega_R}$ their frequency difference. ${R\ast\abs{A^{P/S}}^2}$ correspond to slow pulse-following index modulations, whereas ${R\ast\left[C^{PS}\sin(-\triangle\omega t)\right]}$ represents the index modulation $\triangle\epsilon_R^{\text{osc}}(t)$ oscillating at $\omega_R$ that spectrally shifts the pump through the temporal multiplication ${\triangle\epsilon_R(t)A(t)}$ term \cite{Chen2024,Chen2026}. Growth of the Stokes field is inherently a \emph{spectrotemporal} process: it is shaped by the frequency-dependent Raman gain spectrum [encoded in $R(t)$] and by the time-dependent phonon dynamics -- governed by either the instantaneous intensity $\abs{A(t)}^2$ (steady-state regime) or the integrated pulse energy $E_\text{int}(t)={\int_{-\infty}^t\abs{A(\tau)}^2\diff\tau}$ (transient regime), with ${A(\tau)}$ the electric field \cite{Carman1970,Chen2024}. These effects are all included in $\triangle\epsilon_R^{\text{osc}}(t)$ [Fig.~\ref{fig:long-pulse}(a)], whose bandwidth is determined by the bandwidths of the Raman gain and of the beating amplitude $\mathfrak{F}\left[C^{PS}\right]$. When this oscillatory component is broadband at $\left(-\omega_R\right)$, the generated Stokes spectrum is broader than the pump spectrum, as occurs in single-pulse Raman generation where the Stokes field is temporally confined to part of the pump [top panel of Fig.~\ref{fig:long-pulse}(b)]. When $\triangle\epsilon_R^{\text{osc}}(t)$ is spectrally narrow (extending temporally across the full pulse duration), the pump undergoes an almost rigid spectral shift, yielding a Stokes field that closely reproduces the pump [bottom panel of Fig.~\ref{fig:long-pulse}(b)]: $A^S(t)\sim{\sin(-\omega_Rt)A^P(t)}\sim{\Exp^{-i\omega_Rt}A^P(t)}$. This can arise in transient Stokes generation seeded on the pump's leading edge where long-lived phonons persist across the full pump duration \cite{Chugreev2009}, or in steady-state generation seeded by a long (narrowband) Stokes pulse whose optical beating spans the entire pump. The traditional Raman-gain picture relies on the slow-field and narrowband approximations \cite{Agrawal2013}, which retain only a single value of ${\mathfrak{F}\left[R\right]}$ at a selected ${\triangle\omega}={\omega^S-\omega^P}$, and treats the Raman response as a pre-determined scalar Raman-gain coefficient. Therefore, the Stokes bandwidth in the traditional Raman-gain picture is determined solely by the partial temporal depletion of the pump field, without considering effects of the long-lived transient phonon dynamics [Fig.~\ref{fig:long-pulse}(a)]. As a result of the time-dependent phonon dynamics, the parametric Raman gain suppression is intrinsically time-dependent \cite{Chen2024}, in contrast to the traditional time-independent picture \cite{Bloembergen1964,Duncan1986,Leung1988,Golovchenko1990,Vanholsbeeck2003,Vanholsbeeck2003a}. These considerations restrict the applicability of the Raman gain spectrum to the steady-state regime, where phonon dynamics can be neglected. A more complete description is therefore required to capture the physics in the transient regime, as well as the transitional behavior between the two in weakly-dephasing media or between the impulsive and steady-state regimes in highly-dephasing media. Such a description must incorporate the coupled temporal and spectral dynamics, both of which are embodied in $\triangle\epsilon_R(t)$ within the time-domain framework \cite{Chen2026}.

\clearpage
\bibliography{reference_supplement}